\documentclass[aps,amssymb,amsmath,pra,twocolumn,showpacs,nosuperscriptaddress,longbibliography]{revtex4-2}

\usepackage{graphicx}
\usepackage{dcolumn}
\usepackage{bm}
\usepackage{subfigure}
\usepackage{color}

\usepackage{amssymb,amsmath,amsfonts,latexsym,graphicx,verbatim}

\usepackage[margin=0.5in]{geometry}

\usepackage[english]{babel}
\usepackage{times}
\usepackage{dsfont}
\usepackage{latexsym}
\usepackage{fancyhdr}
\usepackage{dsfont}
\usepackage{dutchcal}
\usepackage{float}
\usepackage{afterpage}
\usepackage{enumitem}
\usepackage{mleftright}

\usepackage{eso-pic, graphicx}

\usepackage{listings}
\usepackage{multirow}
\usepackage{xcolor,colortbl}

\usepackage{bbm}
\usepackage{upgreek}
\usepackage{amsmath}
\usepackage{newtxtext,newtxmath}

\definecolor{Ablue}{rgb}{0.96,0.24,0.00}

\definecolor{Abluetitle}{rgb}{0.,0.24,0.51}

\definecolor{orange}{rgb}{0.96,0.24,0.00}

\definecolor{darkred}{rgb}{0.55, 0.0, 0.0}

\definecolor{darksalmon}{rgb}{0.91, 0.59, 0.48}
\definecolor{maroon}{cmyk}{0,0.87,0.68,0.32}

\definecolor{mustard}{rgb}{1.0, 0.86, 0.35}

\setcitestyle{numbers,square,citesep={,\kern-.24em}}

\definecolor{Gray}{gray}{0.85}
\definecolor{LightCyan}{rgb}{0.88,1,1}
\newcolumntype{a}{$$>$${\columncolor{Gray}}c}
\newcolumntype{b}{$$>$${\columncolor{white}}c}

\usepackage{array}
\newcolumntype{L}[1]{$$>$${\raggedright\let\newline\\\arraybackslash\hspace{0pt}}m{#1}}
\newcolumntype{C}[1]{$$>$${\centering\let\newline\\\arraybackslash\hspace{0pt}}m{#1}}
\newcolumntype{R}[1]{$$>$${\raggedleft\let\newline\\\arraybackslash\hspace{0pt}}m{#1}}

\usepackage[colorlinks=true , citecolor=black,urlcolor=blue]{hyperref}

\newcommand{\xd}{\delta}

\newcommand{\vxe}{\varepsilon}

\newcommand{\xg}{\gamma}
\newcommand{\xt}{\theta}

\newcommand{\xo}{\omega}

\newcommand{\app}{\approx}

\newcommand{\Cs}{{}^{13}\R{C}}

\newcommand{\mC}[0]{\mathcal{C}}
\newcommand{\mI}[0]{\mathcal{I}}

\newcommand{\xD}{\Delta}

\newcommand{\xO}{\Omega}

\newcommand{\mg}[0]{\mathcal{g}}
\newcommand{\mP}[0]{\mathcal{P}}

\newcommand{\sq}[1]{\sqrt{#1}}

\newcommand{\mH}[0]{\mathcal{H}}

\newcommand{\mS}[0]{\mathcal{S}}

\newcommand{\rt}{\rightarrow}

\newcommand{\xrt}[1]{\xrightarrow{#1}}

\newcommand{\beq}{\begin{equation}}
\newcommand{\eeq}{\end{equation}}
                  
\newcommand{\benum}{\begin{enumerate}}
\newcommand{\eenum}{\end{enumerate}}
                    
\newcommand{\bit}{\begin{itemize}}
\newcommand{\eit}{\end{itemize}}

\newcommand{\bea}{\begin{eqnarray}}
\newcommand{\eea}{\end{eqnarray}}

\newcommand{\bev}{\begin{verbatim}}
\newcommand{\eev}{\end{verbatim}}

\newcommand{\non}{\nonumber}

\newcommand{\zt}{\times}

\newcommand{\lb}{\left(}
\newcommand{\rb}{\right)}
\newcommand{\lsb}{\left[}
\newcommand{\rsb}{\right]}

\newcommand{\pll}{\parallel}

\newcommand{\T}[1]{\textbf{#1}}
\newcommand{\I}[1]{\textit{#1}}
\newcommand{\R}[1]{\textrm{#1}}

\newcommand{\zl}[1]{\label{eqn:#1}}
\newcommand{\zr}[1]{Eq.\,(\ref{eqn:#1})}
\newcommand{\zfl}[1]{\protect\label{fig:#1}}
\newcommand{\zfr}[1]{\figurename\,\ref{fig:#1}}

\newcommand{\mat}[4]{\left(\begin{array}{cc}{#1}&{#2}\\
{#3}&{#4}\end{array}\right)}

\newcommand{\ba}{\left\{ \begin{array}{lr}}
\newcommand{\ea}{\end{array}\right.}

\newcommand{\blist}[1]{
 \begin{list}{#1}
 \begin{align}
	 arrow
 \end{align}
 $\checkmark\star
  { \setlength{\itemsep}{3pt}
     \setlength{\parsep}{2pt}
     \setlength{\topsep}{3pt}
     \setlength{\partopsep}{0pt}
     \setlength{\leftmargin}{1em}
     \setlength{\labelwidth}{1em}
     \setlength{\labelsep}{0.5em} } }
\newcommand{\elist}{
  \end{list}  }

\DeclareMathSymbol{\vartheta}{\mathalpha}{letters}{"12}
\DeclareMathSymbol{\theta}{\mathalpha}{letters}{"23}
\DeclareMathSymbol{\phi}{\mathalpha}{letters}{"27}
\DeclareMathSymbol{\varphi}{\mathalpha}{letters}{"1E}

\newcommand{\bef}
{
\begin{figure}[htbp]
\centering
}

\newcommand{\eef}{\end{figure}}

\mleftright
\addto\captionsenglish{\renewcommand{\figurename}{Fig.}}

\newcommand{\beginsupplement}{%
        \setcounter{table}{0}
        \renewcommand{\thetable}{S\arabic{table}}%
        \setcounter{figure}{0}
        \renewcommand{\thefigure}{S\arabic{figure}}%
				
     }

\newcommand{\affA}{Department of Chemistry, University of California, Berkeley, Berkeley, CA 94720, USA.}

\begin{document}
\title{Electron-to-nuclear spectral mapping via \I{``Galton board''} dynamic nuclear polarization}
\author{Arjun Pillai}\affiliation{\affA}
\author{Moniish Elanchezhian}\affiliation{\affA}
\author{Teemu Virtanen}\affiliation{\affA}
\author{Sophie Conti}\affiliation{\affA}
\author{Ashok Ajoy}\email{ashokaj@berkeley.edu}\affiliation{\affA}

\begin{abstract}
We report on a strategy to indirectly readout the spectrum of an electronic spin via polarization transfer to nuclear spins in its local environment. The nuclear spins are far more abundant and have longer lifetimes, allowing repeated polarization accumulation in them. Subsequent nuclear interrogation can reveal information about the electronic spectral density of states. We experimentally demonstrate the method for reading out the ESR spectrum of Nitrogen Vacancy center electrons in diamond via readout of lattice $\Cs$ nuclei. Spin-lock control on the $\Cs$ nuclei yields significantly enhanced signal-to-noise  for the nuclear readout.  Spectrally mapped readout presents operational advantages in being  background-free and immune to crystal orientation and optical scattering. We harness these advantages to demonstrate applications in underwater magnetometry. The physical basis for the \I{“one-to-many”} spectral map is itself intriguing. To uncover its origin, we develop a theoretical model that maps the system dynamics, involving traversal of a cascaded structure of Landau-Zener anti-crossings, to the operation of a tilted \I{“Galton board”}. This work points to new opportunities for \I{“ESR-via-NMR”} in dilute electronic systems, and in hybrid electron-nuclear quantum memories and sensors. 
\end{abstract}

\maketitle

\T{\I{Introduction}} -- The control and readout of electronic spins forms the basis for quantum sensing~\cite{Degen17}. Magnetometry with diamond Nitrogen Vacancy (NV) centers leverages the ability of the NV electronic spin to be optically initialized and read out at room temperature~\cite{Jelezko06,Taylor08}.  Let $\mg(B_0,f)$ denote the NV electronic spectrum at bias field $B_0$ and frequency $f$.  Optical readout (ODMR)~\cite{jelezko04b} maps $\mg$ to fluorescence contrast $S[\mg]$  by exploiting the differential optical contrast between the triplet ground state levels $m_s{=}\{0,\pm 1\}$ originating from state-selective branching in the excited state~\cite{Manson06,Gali08}. Ensemble DC magnetometry then entails measuring a fluorescence change upon a shift in the electronic spectrum, $\xd S{=}S[\mg(B_0+\xd B,f)] - S[\mg(B_0,f)]$ when subject to a probe magnetic field $\xd B$~\cite{Pham11,Barry20}.
	
	Spin-fluorescence mapping, while attractive in many contexts,  suffers from technical limitations that can reduce sensitivity in some operational scenarios.  A typical example is when the N-V axis is misoriented with respect to $\vec{B}_0$~\cite{tetienne12}. State mixing in the excited state then leads to a steep drop in fluorescence and ODMR contrast~\cite{tetienne12}. Even when perfectly aligned, typical fluorescence contrast levels are ${\lesssim} 5\%$ and susceptible to optical backgrounds and readout laser amplitude noise~\cite{Rondin14,Lesage12,Barry20}.  In scenarios requiring quantum sensing in fluidic (or optically dense) media, optical scattering can further reduce contrast. There has been broad interest in alternate strategies that map the NV electronic populations into other parameters (e.g. charge~\cite{Shields15,Siyushev19}) that can be efficiently readout.

\begin{figure}[t]
  \centering
  {\includegraphics[width=0.48\textwidth]{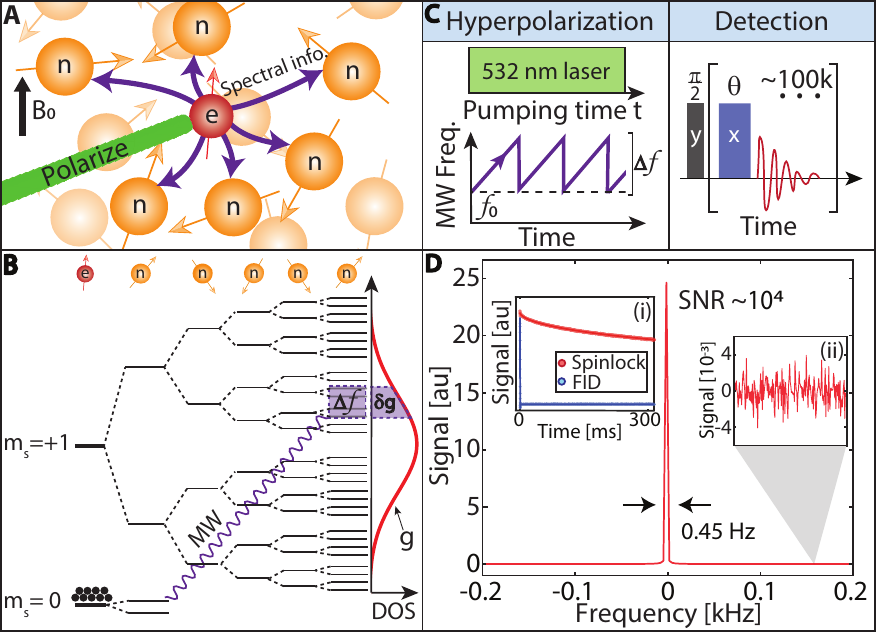}}
  \caption{\T{Spin mapping protocol.} (A) \I{System}. NV center electrons ($e$) surrounded by a network of $\Cs$ nuclei ($n$), which are employed to report the electronic spectrum $\mg$. (B) \I{Principle.} NV${\rt}\Cs$ spectral maps rely on mapping the NV spectral DOS $\xd\mg$ in a narrow $\xD f$ window (purple) to $\Cs$ polarization.  Panel shows the network of energy levels corresponding to an NV center coupled to five $\Cs$ nuclei. Solid red line denotes the resulting NV electronic spectrum $\mg$. Starting with optically initialized populations at $m_s{=}0$ (black balls), chirped microwaves (MWs)~\cite{Ajoy17,Ajoy18} (curly purple line) drive transitions to the $m_s{=}$+1 manifold in the $\xD f$ window (purple). (C) \I{Experimental protocol} consists of DNP applied in narrow $\xD f$ windows (left) and interrogating the $\Cs$ nuclei by spin-locked readout~\cite{Beatrez21} (right), here at 7T, employing ${\sim}$100k pulses. (D) \I{Single shot NMR} spectrum from a representative $\xD f{=}$10MHz window. \I{Inset (i):} time-domain $\Cs$ signal under spin-lock control (red) and free induction decay (blue). For the former, we estimate a rotating frame lifetime $T_2' {\approx} 2$s~\cite{Beatrez21}. \I{Inset (ii):} noise measured in the NMR spectral wing; we estimate a single-shot SNR ${\app}1.25{\zt}10^4$.}
\zfl{fig1}
\end{figure}

\begin{figure*}[t]
  \centering
  {\includegraphics[width=0.98\textwidth]{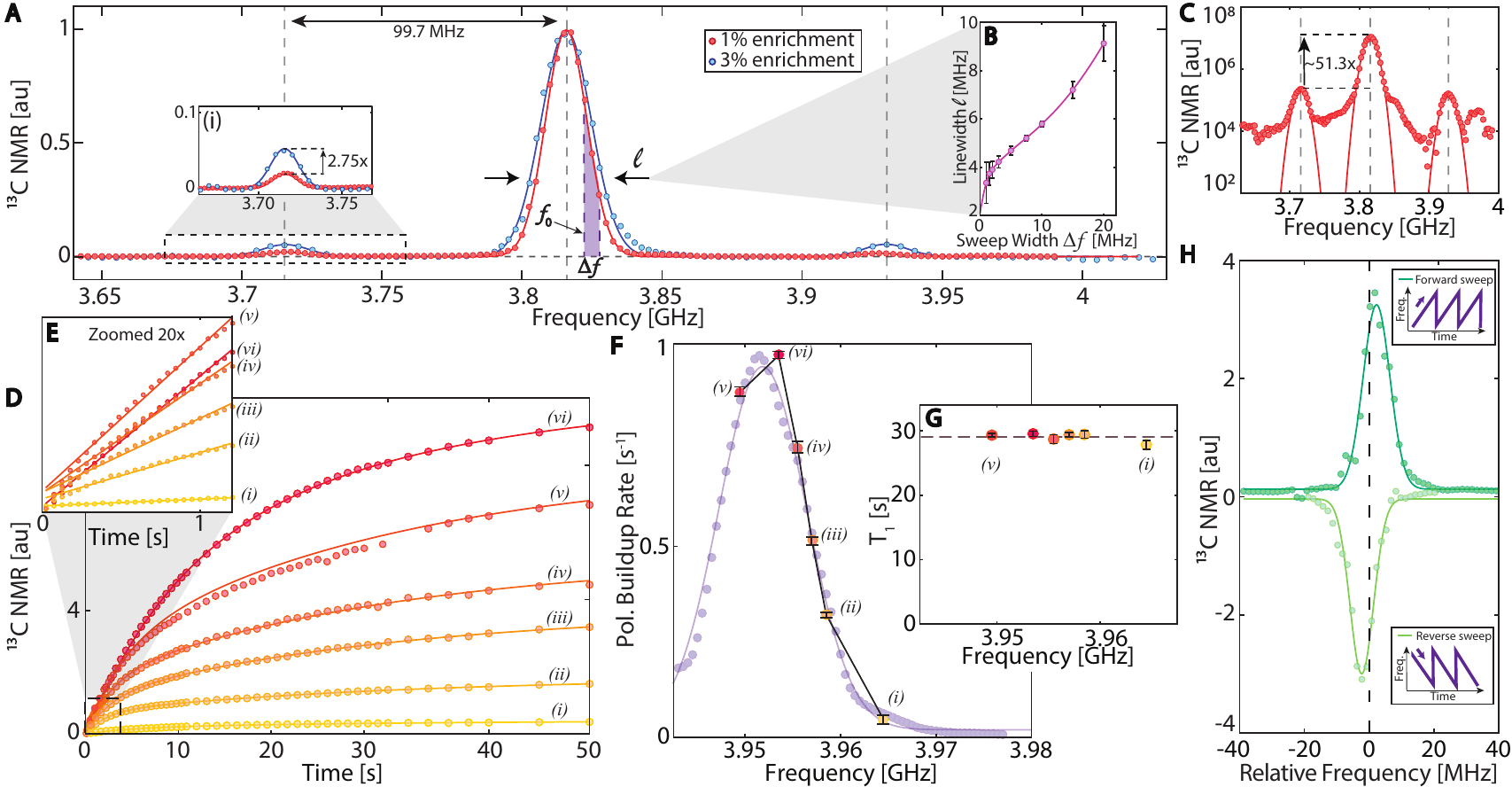}}
  \caption{\T{NV center electronic spectral mapping via $\Cs$ nuclear readout.} (A) \I{NV electronic spectra mapped} to $\Cs$ populations via DNP performed for 20s at $B_0{\approx}$33.6mT in $\xD f{=}$10MHz windows (purple) whose edge is denoted by $f_0$ (see \zfr{fig1}B). Red (blue) data corresponds to 1\% (3\%) $\Cs$-enriched crystals. The four NV families here are aligned at $54^{\circ}$ to $\vec{B}_0$ and overlap in frequency. Solid lines are Gaussian fits. Linewidth $\ell$ increases with enrichment.  Visible are satellites corresponding to NV centers with first-shell $\Cs$ nuclei. \I{Inset (i):} Zoomed view of satellites, offset $-$99.7MHz from the main peak. Normalized spectra show ${\approx}2.75$ greater satellite intensity for the 3\% enriched sample. (B) \I{Linewidth $\ell$ of $\Cs$-mapped NV spectral} peak for a single NV center family for different $\xD f$ window widths. Error bars are calculated from Gaussian fits. Solid line is a spline-fit guide to the eye. Linewidths $\ell$ grow with window size $\xD f$,  and saturate at $\ell{\app} 2$MHz for $\xD f{<}$1MHz. (C) \I{Log scale} representation of data for 1\% enriched sample from (A). Solid lines are Gaussian fits. Satellites are ${\sim}$0.02 the intensity of the main peak. (D) \I{DNP buildup curves} showing $\Cs$ polarization buildup for the different points \I{(i)-(vi)} labeled in (F). Solid lines are biexponential fits. (E) DNP buildup in the small-time limit is approximately linear (solid lines). (F) \I{Extracted buildup rates} (colored points \I{(i)-(vi)}) from the linear fits in (E) with error bars. These rates closely overlay the NV spectrum (lighter purple points), obtained similar to (A). (G) \I{Measured $\Cs$ relaxation times} for the different spectral locations $f_j$ \I{(i)-(vi)} in (D-F) demonstrating that relaxation rates are constant (${\approx}30$s$^{-1}$). (H) \I{$\Cs$-mapped NV spectra obtained under alternate MW sweep directions} with $\xD f$=3MHz. Data here is on a single NV center family. Signal intensities are sweep-sign dependent, but otherwise identical for the entire spectrum. }
\zfl{fig2}
\end{figure*}

Here, we demonstrate an alternate route to reading out the NV electronic spectrum via $\Cs$ nuclear spins in the surrounding environment (\zfr{fig1}A). Our strategy relies on {mapping} the NV spectral density of states (DOS) at frequency $f_0$,  $\xd\mg(B_0,f_0)$,  into $\Cs$ nuclear polarization $P$ (see \zfr{fig1}B),
\beq
\xd\mg(B_0,f_0) = \int_{f_0}^{f_0+\xD f}\mg(B_0,f)df \xrt{\R{map}} P,
\zl{map}
\eeq where $\xrt{\R{map}}$ refers to an approximate map in the limit $\xD f{\rt} 0$. The $\Cs$ nuclei are subsequently inductively readout via RF techniques (NMR) providing certain natural advantages over optical readout~\cite{lv21}: it is crystal orientation independent, background-free, and immune to fluorescence fluctuations due to scattering.

The central focus of this paper is to unravel the mechanism of the NV${\rt}\Cs$ spectral-map in \zr{map}.  Such a \I{one-to-many} spectral map appears challenging at first (\zfr{fig1}A-B): every NV center is surrounded by multiple $\Cs$ nuclei, which are in turn dipole-coupled to each other, and selective $\Cs$ control is difficult. Our approach employs dynamic nuclear polarization (DNP) performed within narrow $\xD f$ spectral windows~\cite{Ajoy17,Ajoy18} (\zfr{fig1}C). We demonstrate that this permits globally mapping the NV DOS $\xd\mg$ into $\Cs$ populations (\zfr{fig1}B).  A key contribution here is uncovering the physical basis behind this process, and hence \zr{map}; we accomplish this by viewing the system dynamics in \zfr{fig1}A, manifesting as traversals through a cascade of Landau-Zener (LZ) anti-crossings, to the analogous operation of a \I{``Galton board”} which ultimately yields analytic solutions.

\begin{figure*}[t]
  \centering
  {\includegraphics[width=0.98\textwidth]{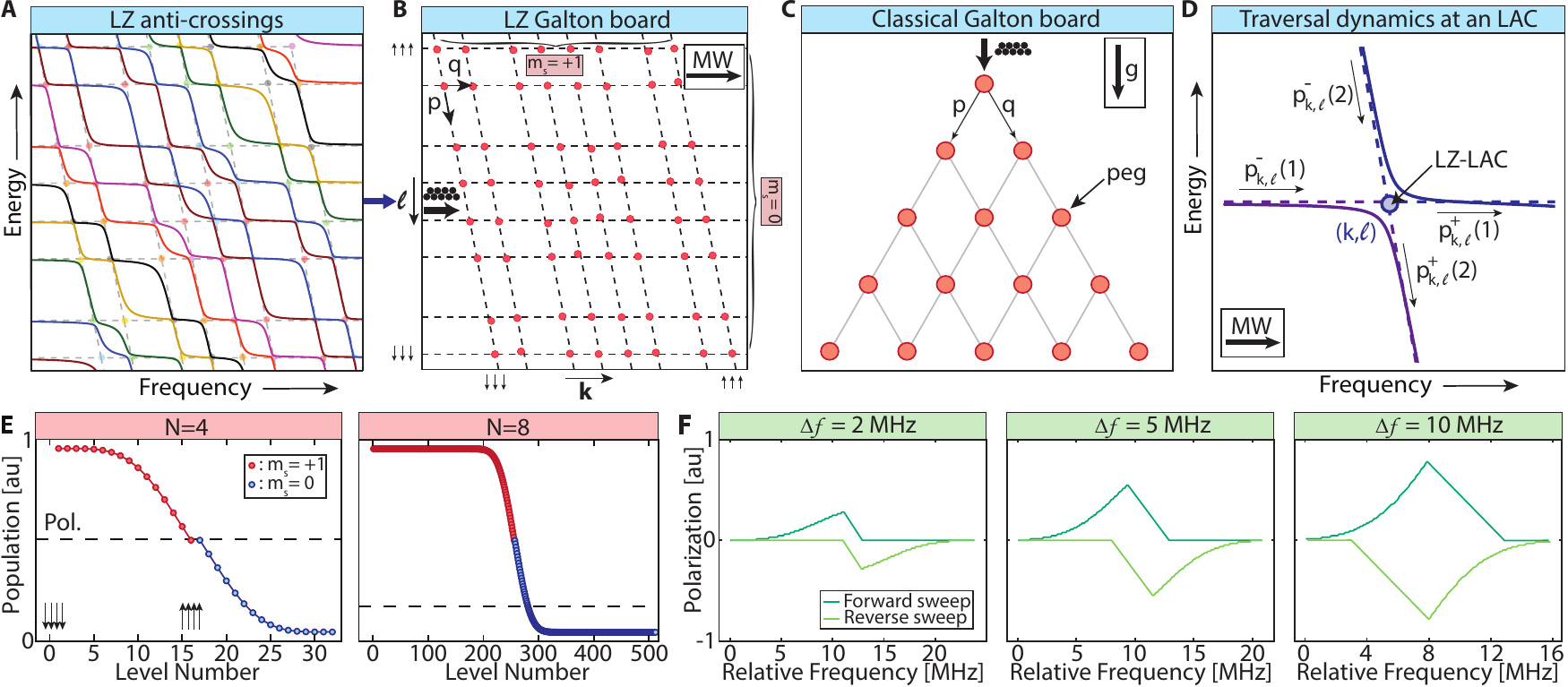}}
  \caption{\T{\I{``Galton board”} nuclear hyperpolarization and DOS mapping.} (A) \I{LZ anti-crossings (LZ-LACs)}. For $N{=}3$, lab frame energy levels of \zfr{fig1}B manifest in the rotating frame as a cascade of LZ anti-crossings (\I{solid lines}) between nuclear states of alternate electronic manifolds. Dashed lines show corresponding crossings and dots highlight LZ-LAC points. (B) \I{Equivalent Galton board}. Isolating LAC points allows us to abstract the system in terms of a Galton board where the crossing points (red) arrange in the tilted checkerboard $\mI_{k,\ell}$ in a 2D plot of energy and frequency (marked $k$ and $\ell$ respectively). The two electronic manifolds are highlighted for clarity. (C) \I{Classical Galton board} consisting of pegs arranged in a pyramidal pattern. Balls striking pegs bounce left or right with probabilities $p$ or $q$ respectively. Driving force is provided by gravity (marked in box). In the analogous LZ Galton board (B), ``balls” are nuclear populations, the ``pegs” are LACs at which populations bifurcate,  and MWs provide the driving force. (D) \I{Transfer matrix description of traversals}. Panel considers a representative section of a larger LAC cascade as in (A), but focused here on only one LAC. Two-component column vectors $\T{p}_{k,\ell}^{\mp}$ denote nuclear populations entering (or leaving) the node $(k,\ell)$. Bifurcation (arrows) of the populations is described by the transfer matrix $\T{T}_{k,\ell}$. (E) \I{Analytical evaluation of population evolution}. Panels show populations $\mP_n$ in the $2^{N}$ numbered nuclear states in the $m_s{=}0$ and $m_s{=}$+1 manifolds (blue and red points respectively) upon a full MW sweep following \zr{eq39} for $N{=}4$ and $N{=}8$ (left and right panels), with $p{=}q{=}0.5$ in (B). Representative nuclear states are marked.  Dashed lines represent the hyperpolarization level. (F) \I{Numerical simulation of DOS mapping}. Panels show hyperpolarization generated from different sized $\xD f$ windows, assuming a Gaussian distribution of energy levels in the $m_s{=}+1$ manifold (see \zfr{fig1}B). Curves track the underlying electronic DOS and become wider with increasing $\xD f$ (as in \zfr{fig2}B). Forward and reverse sweeps yield positive and negative polarization respectively (as in \zfr{fig2}H).}
\zfl{fig3}
\end{figure*}

\T{\I{System}} -- Experiments here are carried out on a single-crystal sample with an ensemble of ${\sim}$1ppm NV centers (\zfr{fig1}A).  The NV spectrum $\mg(B_0,f)$ (red line in \zfr{fig1}B) is constructed out of both homogeneous and inhomogeneous contributions: the former primarily due to interactions with lattice $\Cs$ nuclei, and the latter on account of strain and orientational effects~\cite{Ajoy20}. \zfr{fig1}B, for instance, describes the spectral DOS $\xd\mg$ in window $\xD f$ (purple) arising from an NV center hierarchically coupling to multiple $\Cs$ nuclei~\cite{Ajoy19}.

Our strategy to map the electronic spectrum $\mg$ is outlined in \zfr{fig1}C.  DNP is applied via chirped MWs~\cite{Ajoy17, Ajoy18} in a $\xD f$ window (\zfr{fig1}B) over $\mg(B_0,f_0)$ to transfer the corresponding $\xd\mg$ information to $\Cs$ population differences. This typically involves ${\sim} 10^4$ MW sweeps (left panel of \zfr{fig1}C), during which polarization is accumulated in the $\Cs$ nuclei. \zfr{fig2} demonstrates that the corresponding $\Cs$ polarization closely tracks the electronic DOS $\xd g$ in the $\xD f$ window swept (see \zfr{fig1}B).  A novelty in this work with respect to previous reports~\cite{Alvarez15,Wunderlich17} is the enhanced signal-to-noise (SNR) readout of the $\Cs$ nuclei.  We employ a pulsed spin-locking protocol~\cite{Rhim74,Rhim76} (right panel of \zfr{fig1}C) that allows the $\Cs$ populations to be \I{continuously} interrogated for long periods up to the limit set by the rotating-frame lifetimes,  typically $T_{2}'{>}2$s~\cite{Beatrez21}. It consists of a train of $\xt$-pulses spin-locked with the initial $\Cs$ spin state (\zfr{fig1}C), simultaneously inhibiting dipolar interactions and \I{e}-induced dephasing. Inductive readout (with bandwidth ${\app}$10kHz) occurs after every pulse (red line in \zfr{fig1}C); typical experiments involve ${>}$100k pulses. This yields a $\Cs$ readout SNR that is boosted by ${\gtrsim} 10-10^3$ times~\cite{Beatrez21} with respect to traditional free induction decay (FID) measurements. \zfr{fig1}D shows a representative single-shot $\Cs$ signal obtained using a $\xD f{=}10$MHz window. \zfr{fig1}D\I{(i)} shows the corresponding time-domain signal,  and for comparison, the fast-decaying FID readout. We estimate here a single-shot SNR ${\app}1.25{\zt}10^4$ by comparing to the noise level at the spectral wing (\zfr{fig1}D\I{(ii)}). 

\T{\I{NV${\rt}\Cs$ spectral map}}  -- \zfr{fig2} elucidates results of spectral mapping.  In \zfr{fig2}A, we consider two single crystal samples with $\Cs$ enrichment levels at 1\% (red) and 3\% (blue) respectively. Points here denote the single-shot inductively detected $\Cs$ signal intensity at 7T following \zfr{fig1}C with $\xD f{=}$10MHz (purple window). The high SNR data \I{traces} the underlying NV electronic DOS $\xd\mg(B_0,f)$ with excellent fidelity (solid lines are Gaussian fits), reflecting that the map in \zr{map} is indeed viable. For instance, there is no measurable $\Cs$ signal (to within noise) for $f_0$ where the NV spectral DOS is zero. Intriguingly, however, for a $\xD f$ window even in the wing of the spectrum (purple window in \zfr{fig1}B), the $\Cs$ signal reflects $\xd\mg$.  

We observe two satellites, offset at ${\pm} 99.7$MHz from the spectral center (zoomed in \zfr{fig2}A\I{(i)}). We identify them as arising from NV centers possessing a $\Cs$ nucleus in the first shell~\cite{Rao16}; the high SNR $\Cs$ readout here allows them to be easily discerned despite their low concentration (${\approx} 3\%$ of all NV centers in the 1\% enriched sample). \zfr{fig2}A\I{(i)} shows that while we probe only \I{bulk} $\Cs$ nuclei (hyperfine shifted by ${<}$10kHz), one can unravel $\xd\mg$ information via polarization funneled into them from NV-proximal nuclei. \zfr{fig2}C shows data for the 1\% enriched sample in a logarithmic scale; the ratio of the satellite and main peak intensities (${\app}$0.02) approximately follows the expected 3.3\% concentration. Returning to \zfr{fig2}A, it is evident that increasing $\Cs$ enrichment yields a larger electronic spectral broadening~\cite{Ajoy19relax} and increased (relative) satellite intensity, both as theoretically predicted. \zfr{fig2}B describes the measured spectral linewidth $\ell$ (arrows in \zfr{fig2}A), as a function of the sweep window $\xD f$ employed. It increases approximately linearly with $\xD f$, as expected from \zr{map}, and saturates at $\ell{\app}2$MHz, which correlates with the linewidth due to strain-broadening~\cite{Acosta09}. Error bars here are obtained from the Gaussian fits.
	
	Some key features of the spectra in \zfr{fig2}A are worth mentioning. Here, the NV center families have their NV axes at 54$^{\circ}$ to the bias field $\vec{B}_0$: a regime of strong misorientation where ODMR suffers significant (${\gtrsim}70\%$) contrast loss~\cite{Rondin14}. RF readout here is background free, and spin-locking provides a significantly higher SNR compared to previous reports~\cite{Parker19,Pagliero17,Alvarez15,Wunderlich17}. As opposed to solid-effect DNP~\cite{Hovav10} or that obtained at the NV ESLAC~\cite{Fischer13,Pagliero17}, the sign of the $\Cs$ NMR signals obtained via \zfr{fig1}C is identical for the entire spectrum in \zfr{fig2}A~\cite{Zangara18}. This makes unraveling the electronic spectra feasible even when they are inhomogeneously broadened (see \zfr{fig4}). 
	
What is the physical origin of a higher $\Cs$ polarization for a spectral point $f$ at the peak in \zfr{fig2}A versus at the wing? Insight is provided by \zfr{fig2}D-E which consider how the \I{growth} of nuclear polarization relates to the electronic DOS. We study polarization buildup curves for representative points $f_j$ on the spectrum $\mg(B_0,f_j)$ (marked \I{(i)-(vi)} in \zfr{fig2}F) for the natural abundance (1\%) sample.  Indeed, the points in \zfr{fig2}A reflect a slice of this data at $t{=}20$s.  Polarization buildup reflects an interplay between spin injection into directly hyperfine coupled $\Cs$ nuclei, spin diffusion, and $\Cs$ relaxation. To a good approximation, $\Cs$ relaxation rates are determined by interactions with paramagnetic impurities (P1 centers)~\cite{Reynhardt98,Ajoy19relax}, and are independent of the spectral location $f_j$ being probed. This is shown in \zfr{fig2}G, where we find $T_{1n}{{\approx}}30$s (dashed line) independent of $f_j$ (see ~\cite{SOM}). Similarly, spin diffusion rates, driven by internuclear interactions, can be considered to be independent of $f_j$~\cite{Abragam61}.  One expects then that data in \zfr{fig2}A arises from differences in polarization injection \I{rates} conditioned on $f_j$, this in turn produces the frequency-dependent polarization levels and yields \zr{map}.  Experiments in \zfr{fig2}D-F confirm this intuition. We restrict attention to short-time (${<}$1s) buildup (zoomed in \zfr{fig2}E) and linearize the corresponding polarization growth curves (solid lines) to extract the injection rate. Normalizing and overlaying these values (colored points) on the normalized $\Cs$-interrogated NV electronic spectra in \zfr{fig2}F (lighter purple points) shows that the polarization injection rates produce the signal differences that tracks $\xd \mg(B_0,f_j)$.

Finally, \zfr{fig2}H illustrates the result of an experiment similar to \zfr{fig2}A (with $\xD f{=}$3MHz), but focused on a single NV center family and employing alternate MW sweeps conditions~\cite{Ajoy17,Zangara18} (low-to-high or high-to-low frequency). Both cases in \zfr{fig2}H reflect the electronic spectrum, but with an opposite sign. There is a small observable shift ($\app$2.3MHz) from the spectral center.

\T{\I{``Galton board” hyperpolarization}} --  Experiments in \zfr{fig2} give rise to the intriguing question: why is the map in \zr{map}, even if just approximate, possible at all? This can be recast to asking why DNP in a $\xD f$ window in the \I{wing} of the spectrum (as in the purple window in \zfr{fig2}A) should produce a signal tracking $\xd\mg$.  Modeling polarization transfer at the spectral wing requires the ability to solve dynamics for an electron coupled to multiple ($N$) nuclei (as in \zfr{fig1}A-B) under a MW sweep. 
This requires extending analytic descriptions of DNP mechanisms beyond commonly considered $e$-$n$ or $e$-$e$-$n$ systems~\cite{Hovav10,Thurber12,Zangara18,Hovav12}. Here we develop an approach to make such solutions tractable at large $N$. A companion paper~\cite{Elanchezhian21} elucidates this theory in greater detail; here we present its salient aspects and connections to experiments.

The instantaneous Hamiltonian of the \I{e-n} system at a particular frequency $f_0$ along the MW sweep is, 
\beq
\mH (f_0) = (\xD - f_0)S_z^2 + \xg_e B_0 S_z + \xO_e S_x + \sum_{j=1}^{N}\lsb \xo_j^{(0)}\mP_0I_{zj} + \xo_j^{(1)}\mP_1I_{z'j}\rsb\non
\zl{eq4}
\eeq
where we restrict attention to the $m_s{=}\{0,+1\}$ manifold,  ignore internuclear interactions, and where $\xD$=2.87GHz is the NV center zero field splitting, $\xO_e$ is the electronic Rabi frequency, and $\mP_0{=}\mat{1}{0}{0}{0}\otimes\T{1}_N\:,\: \mP_1{=}\mat{0}{0}{0}{1}\otimes\T{1}_N$ are projection operators. The eigenenergies of $\mH (f_0)$ manifest as a cascade of LZ-LACs, the number of which scale exponentially with $N$. \zfr{fig3}A shows this for a representative example of $N{=}3$ and positive hyperfine couplings $A_j^{\pll}{>0}$. 

Modeling experiments in \zfr{fig2}A requires a machinery to track the nuclear populations as this LZ cascade is traversed. We first extract the $2^{2N}$ LAC points to a 2D checkerboard $\mI_{k,l}$ (\zfr{fig3}B) of energy and frequency (axes labeled $k$ and $\ell$), and identify their corresponding energy gaps as $\vxe_{k,\ell}$. Upon optical pumping, the initial nuclear populations are equally distributed among nuclear states in the $m_s{=}0$ manifold. Then under the MW sweep, the nuclear populations traverse through $\mI_{k,l}$ and redistribute down or right at every LAC (\zfr{fig3}B). These traversals can be viewed as analogous to the operation of a classical Galton board~\cite{Bouwmeester99,Chernov07} (\zfr{fig3}C) in which balls fall through a system of pegs under gravity, encountering which they bounce left or right. In a similar manner, the LACs in \zfr{fig3}B form the “pegs”, the “balls” are the nuclear populations, and the swept MWs provide the driving force. The probability of population bifurcation ``right'' at each LAC is conditioned on the size of the energy gap $\vxe_{k,\ell}$, and given by the tunneling probability $\eta_{k,\ell}{=}\exp(-\vxe_{k,\ell}^2/ \dot{f_0})$, where $\dot{f_0}$ is the MW sweep rate. Downward bifurcation then has probability $1-\eta_{k,\ell}$. The energy gaps $\vxe_{k,2^{N}-k+1}{\app}\xO_e$ are the largest on the checkerboard, making traversals through them approximately adiabatic.

We assume that the LACs in \zfr{fig3}B are hit sequentially due to the \I{tilted} nature of the LZ Galton board, and that despite continuous action of the laser, electronic repolarization happens far away from the LACs. If $\T{p}_{k,\ell}^{-}$ and $\T{p}_{k,\ell}^{+}$ are two-element column vectors denoting the nuclear populations \I{entering} or \I{leaving} the $(k,\ell)$ LAC (\zfr{fig3}D), then
$
 \T{p}_{k,\ell}^{+}{=}\T{T}_{k,\ell}\: \T{p}_{k,\ell}^{-},
$
where
$
\T{T}_{k,\ell} =\mat{\eta_{k,\ell}}{\left(1- \eta_{k,\ell}\right)}{\left(1- \eta_{k,\ell}\right)}{\eta_{k,\ell}}
$ is a transfer matrix~\cite{Katsidis02,Troparevsky10} describing the bifurcation of populations analogous to a Galton board. Traversals through LACs as in \zfr{fig3}A can then be \I{recursively} solved as, 
$
\T{p}_{k,\ell}^{+}{=}\T{T}_{k,\ell} \T{M}_{r} \T{p}_{k-1,\ell}^{+}+\T{T}_{k,\ell} \T{M}_{d} \T{p}_{k,\ell-1}^{+},
$
where the operators,
$
\T{M}_{d}{=}\mat{0}{0}{0}{1}\ \R{and}\ \T{M}_{r}{=}\mat{1}{0}{0}{0}\:,
$
describe ``walks” down and right through the $\mI_{k,\ell}$ board in \zfr{fig3}B. Ultimately,  one can quantify the traversal probability between two points with coordinates $(k_i, \ell_i)$ and $(k_f, \ell_f)$ in $\mI_{k,\ell}$ as~\cite{Elanchezhian21},
\beq
\mP[(k_i,\ell_i){\rt} (k_f,\ell_f)] = \sum_{\{v\}\in L_p}\lb \prod_{j=1}^{n}\mC_{v_{j}}\rb ,
\zl{eq28}
\eeq
where $L_p{=}\binom{L}{k_f-k_i}$ represents the total number of paths $\T{v}$ involved, defined by the nearest neighbor vertices such that $\T{v}{=}\{v_j\}{=} \{(k_1,\ell_1), (k_2,\ell_2),\cdots, (k_{n},\ell_{n})\}$, and the coefficients, 
\beq
\mC_{v_{j}}=\left\{
                \begin{array}{ll}
                  \eta_{k_j,\ell_j}& \text{if $k_{j-1} = k_{j+1}$ or $\ell_{j-1} = \ell_{j+1}$}\\
                  (1-\eta_{k_j,\ell_j})& \text{if $k_{j-1} \neq k_{j+1}$ and $\ell_{j-1} \neq \ell_{j+1}$}\\
                \end{array}
              \right.
\zl{eq27} .
\eeq

After the action of the laser that resets the NV to $m_s{=}0$, the population of any nuclear state $\mS$ is,
$
\mP_{n} {=} {p}_{2^{N},\ell_{n}}^{+}(1) + {p}_{k_n,2^{N}}^{+}(2),
$
where $n \in 1{\cdots}2^{N}$ indexes $\mS$ in a Hamming ordering. Consequently, the nuclear hyperpolarization $P$ in \zr{map} and measured in experiments in \zfr{fig2}A takes the form,
\beq
P=\sum_{n=1}^{2^{N-1}}{\mP}_n - \sum_{n=2^{N-1}+1}^{2^{N}}{\mP}_n,
\zl{eq29}
\eeq 
for $A^{\pll}{>}0$. The alternate case of $A^{\pll}{<}0$ is shown to produce no hyperpolarization in the $m_s{=}+1$ manifold~\cite{Elanchezhian21}. 

Drawing the analogy to the classical Galton board to yield the solution procedure in \zr{eq28}-\zr{eq29} for the coupled \I{e-}$(n)^N$ system in \zfr{fig1}A is a key result of this work. It permits an analytical (or numerical) means to evaluate  traversals through the full $\mI_{k,\ell}$ checkerboard in \zfr{fig3}B, or in narrow $\xD f$ window through it (as in \zfr{fig2}A). Consider first the solution of the \I{full} sweep through $\mI_{k,\ell}$. Assuming the probability of redistribution down and right to be $p$ and $q$ respectively at every LAC and that the large energy gaps yield adiabatic traversals ($\eta_{k,2^{N}-k+1}{=}0$), assumptions that do not alter the essential physics of the problem, the probability of ending in the nuclear state $n {\in}\mS$ after a \I{full} MW sweep can be \I{analytically} written as~\cite{Elanchezhian21}:
\begin{widetext}
\bea
\mP_{n}=\frac{1}{2^{N}}\Bigg(\sum_{\ell=1}^{2^{N}-1}\Bigg[{(n{-}2){+}(2^{N}{-}\ell{-}1)\choose n{-}2}p^{2^{N}{-}\ell}q^{n{-}2}{+}{(2^{N}{-}n{-}\ell)+(2^{N}{-}2)\choose 2^{N}{-}n{-}\ell}p^{2^{N}{-}n{-}\ell}q^{2^{N}{-}1}\Bigg]+c(n)\Bigg)\zl{eq39},
\eea
\end{widetext}
where $\cdot\choose{\cdot}$ represents the combinatoric operator and $c(n)$ is a constant equal to 1 when $n{=}1$ and 0 otherwise.  \zr{eq39} alludes to the binomial nature of Galton board traversals~\cite{Lue93} and provides insight into the physics of the polarization transfer process.  Solutions to $N{=}4$ and $N{=}8$ are displayed in \zfr{fig3}E, showing populations of the numbered nuclear states in the $m_s{=}0$ and $m_s{=}+1$ manifolds (blue and red points respectively). Representative nuclear states are marked.  As is evident, traversal through the Galton board in \zfr{fig3}B is ``\I{biased}", resulting in the hyperpolarization $P$ denoted by the horizontal dashed lines in \zfr{fig3}E. 

In a similar manner,  one can carry out calculations of traversal through a narrow $\xD f$ window on the LZ Galton board (\zfr{fig2}A). While an analytic solution is tractable, it is unwieldy and here we resort to a numerical solution assuming an exemplary Gaussian distribution of energy levels $\mg$ (with width 13.5MHz) in the $m_s{=}+1$ manifold as in \zfr{fig1}B. The edge of the window ($f_0$) is swept in an analogous fashion to the experiments in \zfr{fig2}A, and nuclear populations are calculated following \zr{eq28}-\zr{eq29}. \zfr{fig3}F demonstrates the result for varying $\xD f$ and under alternate sweep conditions. We observe that the polarization levels $P$ approximately track the Gaussian DOS $\xd\mg$, qualitatively matching the experimental results in \zfr{fig2}.  Intuitively, this is because the Galton board operation ``moves'' nuclear populations following the underlying distribution $\xd\mg$. Larger $\xD f$ windows (panels left-to-right in \zfr{fig3}F) yield a broader spectrum,  and opposite sweeps produce an alternate hyperpolarization sign (\zfr{fig3}F). These match the experimental observations in \zfr{fig2}B and \zfr{fig2}H.

\begin{figure}[t]
\centering
{\includegraphics[width=0.49\textwidth]{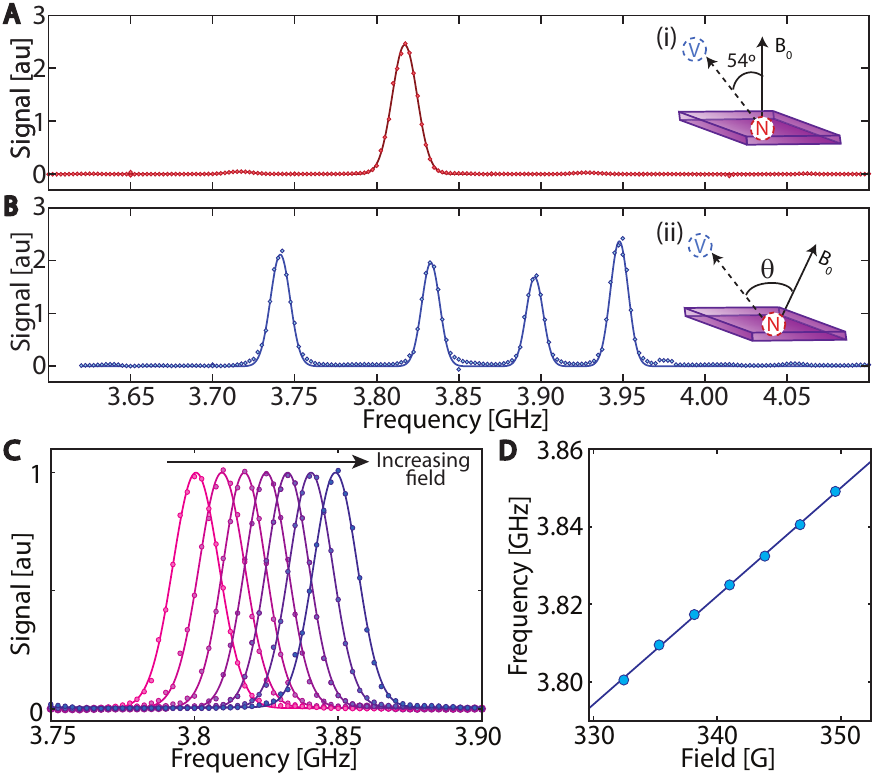}}
\caption{\T{Underwater DC magnetometry via $\Cs$ readout} performed with the sample submerged in ${\app}$4.9mL water. (A-B) \I{NV spectral maps of misoriented crystals} for when the NV centers are at (A) $\xt{=}54^{\circ}$ and (B) arbitrarily oriented to the bias field $B_0{=}33.6$mT (shown in insets (i) and (ii) respectively). (C) \I{DC magnetometry. }Nuclear mapped NV spectra for A(i) with different applied magnetic fields $\xd B$ collinear with $B_0$. (D) Scaling of spectral center frequency with field. }
\zfl{fig4}
\end{figure}

\T{\I{Outlook and conclusions}} — This work opens many interesting future directions. \zfr{fig2}A suggests the possibility to perform \I{“ESR-via-NMR”}, portending avenues to probe electronic spins via nuclei coupled to them. We envision applications in dilute electronic systems or photopolarizable radicals~\cite{Rugg19}, where ESR spectra may be inaccessible, but where information can be instead relayed to more abundant, longer-lived nuclear spins to be readout. Such \I{“one-to-many”} spectral maps also suggest applications in quantum memories~\cite{Bradley19}.

From a technological perspective, we envision applications in RF ($\Cs$) interrogated NV-center magnetometry, \I{without} the use of a MW cavity~\cite{Eisenach21}. As opposed to optical NV sensors, this can permit DC magnetometers that function in turbid or scattering media and with arbitrarily oriented crystals.  \zfr{fig4} demonstates a proof-of-concept, employing NV${\rt}\Cs$ maps for bulk DC magnetometry \I{underwater}, a regime with several applications (e.g. undersea magnetic anomaly detection~\cite{Ge20}), but where current quantum sensor technologies are not viable.  \zfr{fig4}A-B shows NV spectra for samples placed under ${\app}4.9$mL of water,  corresponding to about 2000-fold the volume of the sample.  Here the four NV axes are identically aligned at 54$^{\circ}$ to the bias field (\zfr{fig4}A) and arbitrarily oriented (\zfr{fig4}B). We clarify that only the sample (and not the excitation and detection apparatus) is submerged, but these experiments suggest that fully underwater quantum sensor magnetometers are feasible.

DC magnetometry can be carried out by monitoring the NV spectral shift under an applied probe field $\xd B$. \zfr{fig4}C-D shows the resulting individual $\Cs$-interrogated spectra, from where we estimate an underwater DC sensitivity ${\app} 363\R{nT}/\sq{\R{Hz}}$. Sensitivity is governed by the NMR readout SNR,  which is still limited here due to the relatively low NMR coil filling-factor (${\app}$0.004), $\Cs$ hyperpolarization level ($\app$0.2\%) and small readout time (0.3s) with respect to the dead-time. These factors can be considerably improved with advancements to the experimental apparatus (see Ref.~\cite{Beatrez21}); we estimate a sensitivity improvement by four orders of magnitude is feasible. \zfr{fig4}B also suggests underwater vector magnetometry exploiting the four NV center families simultaneously~\cite{Schloss18}. 

In conclusion, we have demonstrated a technique to readout electronic spectra by \I{mapping} them to spin population differences in surrounding nuclear spins via DNP. We applied this to mapping the NV center spectrum $\mg$ into the $\Cs$ spin polarization levels. We showed that this physically originates from the DNP process stemming from traversals of cascaded structure of LZ anti-crossings. In an analogy to the operation of a ``Galton board”, we showed that these traversals can be analytically solved, yielding a means to extract the resulting hyperpolarization levels in the $N$ nuclear spins coupled to the NV center. Finally, we demonstrated practical applications of the method for underwater bulk-diamond DC magnetometry. We envision applications of the spectral mapping technique introduced here to a broader class of hybrid quantum systems. 

We acknowledge discussions with J. Reimer, S. Bhave, P. Zangara, C. Meriles and D. Suter. This work is funded by ONR under contract N00014-20-1-2806.  

\bibliography{Galton_short_FINAL2.bbl}

\pagebreak

\clearpage
\onecolumngrid
\begin{center}
\textbf{\large{\textit{Supplementary Information} \\ \smallskip
Electron-to-nuclear spectral mapping via \I{``Galton board''} dynamic nuclear polarization}}\\
\hfill \break
\smallskip
Arjun Pillai, Moniish Elanchezhian, Teemu Virtanen, Sophie Conti and Ashok Ajoy\\
\emph{{\small Department of Chemistry, University of California, Berkeley, Berkeley, CA 94720, USA.}}

\end{center}

\twocolumngrid

\beginsupplement

\begin{figure}[t]
  \centering
  {\includegraphics[width = 0.5\textwidth]{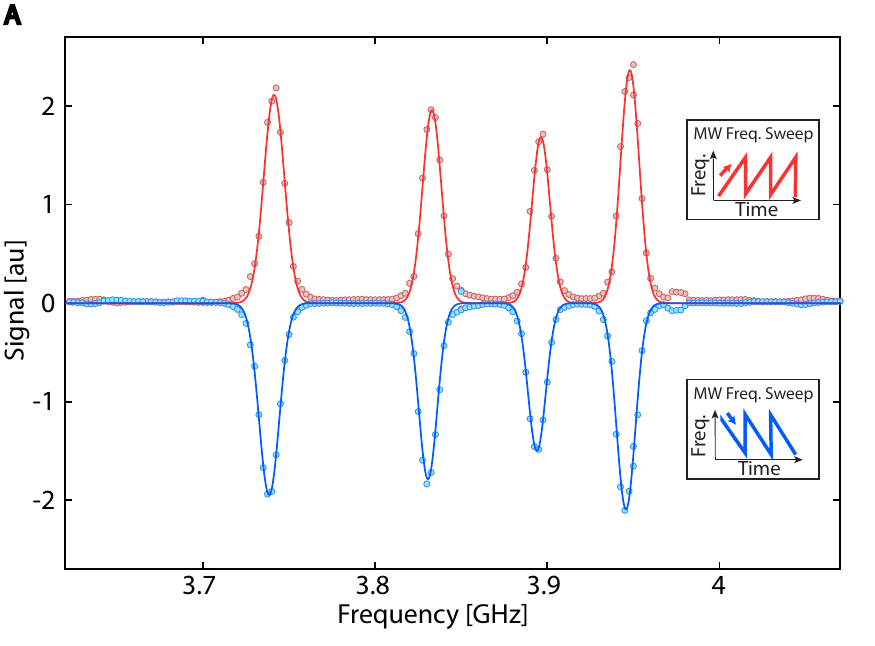}}
  \caption{\T{NV spectral mapping via $\Cs$ readout. }  Data here is for four NV center families, extending \zfr{fig4}B of the main paper. Red (blue) curves here represent $\Cs$ hyperpolarization signals obtained under low-to-high (high-to-low) frequency sweeps.}
\zfl{figS1}
\end{figure}

\begin{figure}[t]
  \centering
  {\includegraphics[width = 0.5\textwidth]{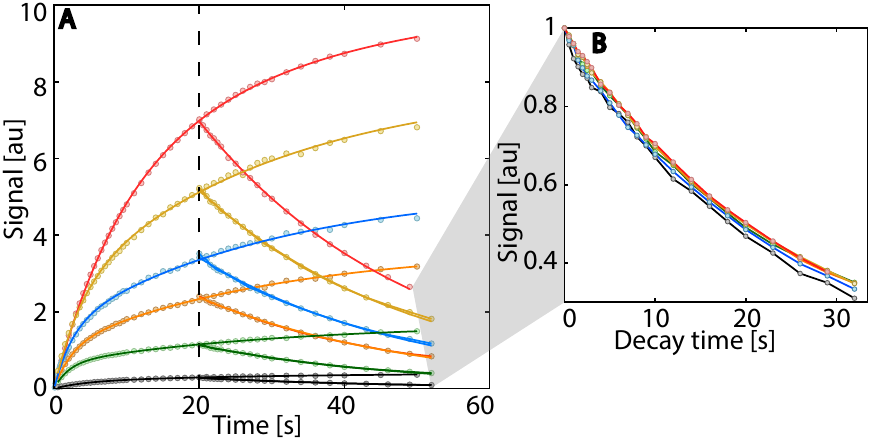}}
  \caption{\T{Polarization injection and relaxation dynamics. } Raw data corresponding to \zfr{fig2}G of the main paper. (A) Here $\Cs$ relaxation rates measured for different $\xD f{=}20$MHz windows are measured by allowing the sample to relax in the polarizing field subsequent to hyperpolarization for $t{=}20$s. Both the polarization buildup curves (also shown in \zfr{fig2}D-E) of the main paper and the corresponding relaxation decays are shown. (B) Overlaying the normalized relaxation curves in (A) shows close overlap, indicating that the relaxation rates are independent of the $f_j$ spectral location chosen.  }
\zfl{figS2}
\end{figure}

\section{Extended Data}
 \vspace{-1mm}
Extended data for \zfr{fig2} and \zfr{fig4} of the main paper are shown in \zfr{figS1} and \zfr{figS2}. First, \zfr{figS1} considers data corresponding to \zfr{fig4}B of the main paper for an arbitrarily oriented diamond crystal showing the four NV center families. We note that for each family, the $\Cs$-obtained NV ESR spectrum is of the same sign, which is opposite for the two MW sweep directions. This highlights the main property that makes it easy to uncover the electronic spectrum in spite of the spectra having inhomogenously broadened components.

\zfr{figS2}A shows the raw data corresponding to \zfr{fig2}G of the main paper. Here, for different parts of the ESR spectrum (shown in \zfr{fig2}F) of the paper, we determine the polarization buildup characteristics at a bias field $B_0$ (also shown in \zfr{fig2}D). After 20s of polarization injection (dashed line), the $\Cs$ spins are then allowed to relax at the same bias field $B_0$. This relaxation data is normalized and shown in \zfr{figS2}B, from where it is evident that the relaxation characteristics are independent of the NV ESR spectral location being originally polarized. The extracted time constants of this decay were plotted in \zfr{fig2}G of the main paper.

\end{document}